\documentclass[
preprint,
amsmath,amssymb,aps,
]{revtex4-1}

\usepackage{graphicx}
\usepackage{dcolumn}
\usepackage{bm}

\begin{document}


\title{Seesaw Scale and CP Phases in a Minimal Model of Leptogenesis}

\author{Kim Siyeon}%
 \email{siyeon@cau.ac.kr}
 \affiliation{Department of Physics, Chung-Ang University, Seoul 06974 Korea}

\date{November 14, 2016}

\begin{abstract}
The seesaw mechanism to derive the light masses of left-handed neutrinos using heavy masses of right-handed neutrinos gives rise to a connection between low-energy measurables and GUT-scale mechanism. We expresses the neutrino mixing angles in terms of a single variable $\sin\theta_{13}$, whose size was measured recently. The lepton asymmetry from heavy neutrinos via Yukawa coupling is described by CP phases in both Dirac and Majorana type. It is shown that the seesaw scale relevant to the lepton asymmetry can be constrained by CP phase in this minimal model.
\begin{description}
\item[PACS numbers]
11.30.Fs, 14.60.Pq, 14.60.St
\item[Keywords] leptogenesis, neutrino mass, CP asymmetry, seesaw mechanism
\end{description}
\end{abstract}

\maketitle

\section{\label{sec:level1}Introduction}

The transformation between three active neutrinos of Standard Model(SM) and massive neutrinos is almost understood from the measurements of three mixing angles except a CP phase \cite{atm}\cite{SNO}\cite{An:2012eh}\cite{Ahn:2012nd}. The current global analysis presents the following best fits: $\sin^2\theta_{12}=3.08\times 10^{-1}, ~\sin^2\theta_{23}=4.37\times 10^{-1},$ and $\sin^2\theta_{13}=2.34\times 10^{-2}$ for normal hierarchy mass ordering(NH), i.e., assuming $m_1 < m_2 < m_3$ \cite{Beringer:1900zz}\cite{Capozzi:2013csa}\cite{Tortola:2012te}. The current knowledge on neutrino masses is still nothing but the mass-squared differences, $m_3^2-m_1^2=2.43\times 10^{-3}\mathrm{eV}^2$ and $m_2^2-m_1^2=7.54\times 10^{-5}\mathrm{eV}^2$ for atmospheric neutrinos and solar neutrinos, respectively. Recent measurements of the third mixing angle $\theta_{13}$ have been naturally followed by search of the CP phase in Pontecorvo-Maki-Nakagawa-Sakata(PMNS) .
The CP conservation at $3\sigma$ confidence level(CL) has been excluded by the result of T2K  \cite{Escudero:2016odp}. A next-generation oscillation experiment is rushing to narrow down the range of $\delta$, the CP phase of PMNS matrix \cite{Acciarri:2015uup}.

Leptogenesis is a theory regarded as an explanation of the Baryon-antibaryon Asymmetry in Universe(BAU) \cite{lep}\cite{Harvey:1990qw}\cite{Kolb:qa}, in which an indirect test is possible. If the decays of heavy Majorana neutrinos via Yukawa couplings are the sources of leptonic CP violation and the heavy neutrinos are the elements in seesaw mechanism \cite{Barger:2003gt}\cite{Luty:un}, some parameters can be tested phenomenologically \cite{Endoh:2002wm}. It is worthwhile to examine the implication of the sizes of $\theta_{13}$ and $\delta$ in a canonical leptogenesis model. Seesaw mechanism is a top-down approach in which the light neutrino masses are obtained through Grand Unified Theory(GUT)-scale mechanism \cite{Gell-Mann:vs}. In this work, bottom-up approach is taken to probe the lepton asymmetry of heavy-neutrino decays in high-energy scale by using the masses and the mixing angles of light neutrinos in low energy.

In order to see the effect from the definite size of $\theta_{13}$ on the high-energy asymmetry, all the elements of PMNS matrix are expressed in terms of a single variable $\sin\theta_{13}$ except CP phase. One of the minimal choices is made for seesaw mechanism, which is that Yukawa matrix is constructed only with two right-handed neutrinos.
We found the way to express $3 \times 2$ Dirac mass matrix by matching the light neutrino mass matrix obtained from the seesaw mechanism to the low energy neutrino mass matrix in weak interaction basis. Our previous work on the seesaw mechanism in bottom-up approach was also based on $3 \times 2$ structure for Dirac mass, although a texture zero in matrix was forced \cite{Lee:2005cda}. This work is outlined as follows: The neutrino mass matrix is constructed with masses and unitary transformation in Section II. We derive a mass matrix using the seesaw mechanism and CP asymmetry from the decay of right-handed neutrino via Yukawa coupling in Section III. In Section IV, We examine the dependency of the lepton asymmetry on low-energy parameters. It is worthwhile to discuss the correlation of the lepton asymmetry with CP phases in PMNS matrix.  In conclusion, we summarize the relation of the lepton asymmetry with low-energy measurables and the possibility to narrow down the models of seesaw mechanism.

\section{Low-energy constraint}

The PMNS mixing matrix for 3 generations of Majorana neutrinos
is given by
    \begin{eqnarray}
    \tilde{U} &=& R\left(\theta_{23}\right)
          R\left(\theta_{13},\delta\right)
          R\left(\theta_{12}\right)P(\varphi_2, \varphi_3)
    \label{fulltrans}
    \end{eqnarray}
where each $R$ is a rotation matrix with a mixing angle $\theta_{ij}$ between $i$-th and $j$-th generations. According to the standard parametrization, the Dirac phase is combined with $\theta_{13}$ as $R\left(\theta_{13},\delta\right)$ in the PMNS matrix. A diagonal phase transformation $P(\varphi_2, \varphi_3)$ is given by Diag$\left(1,e^{i\varphi_2/2},e^{i\varphi_3/2}\right)$. The Majorana phases $\varphi_2$ and $\varphi_3$ can be a part of the mass matrix of light neutrinos in the following way:
    \begin{equation}
    M_{\nu} = U Diag(m_1,\check{m}_2,\check{m}_3) U^T,
    \label{umu}
    \end{equation}
where $U \equiv \tilde{U}P^{-1}$, $\check{m}_2 \equiv
m_2e^{i\varphi_2}$ and $\check{m}_3 \equiv m_3e^{i\varphi_3}$. The standard parametrization for Cabibbo-Kobayashi-Maskawa(CKM) matrix can be taken for transformation matrix $U$ in the PMNS such as
\begin{widetext}
    \begin{eqnarray} U=
        \left(\begin{array}{ccc}
            c_{12}c_{13} & s_{12}c_{13} & s_{13}e^{-i\delta}\\
            -s_{12}c_{23}-c_{12}s_{23}s_{13}e^{i\delta} &
            c_{12}c_{23}-s_{12}s_{23}s_{13}e^{i\delta} &
            s_{23}c_{13} \\
            s_{12}s_{23}-c_{12}c_{23}s_{13}e^{i\delta} &
            -c_{12}s_{23}-s_{12}c_{23}s_{13}e^{i\delta} &
            c_{23}c_{13}
        \end{array}\right),\label{ckm}
    \end{eqnarray}
\end{widetext}
where $s_{ij}$ and $c_{ij}$ denotes $\sin{\theta_{ij}}$ and
$\cos{\theta_{ij}}$.

The measurement of $\theta_{13}$ from recent reactor anti-neutrino detection allows the following conditions:
	\begin{eqnarray}
	s_{12}=\frac{1}{\sqrt{2}}-s_{13}, \label{s12}\\
	s_{23}=\frac{1}{\sqrt{2}}-s_{13}^2. \label{s23}
	\end{eqnarray}
Such deviation from bi-maximal mixing scheme was introduced in a number of models. By substituting Eq.(\ref{s12}) and Eq.(\ref{s23}) into Eq.(\ref{ckm}) the neutrino mixing matrix in the leading order can be written as;
\begin{eqnarray}
    U &\approx&
	 \left(\begin{array}{ccc}
    \frac{1}{\sqrt{2}} & \frac{1}{\sqrt{2}} & 0 \\
	-\frac{1}{2} & \frac{1}{2} & \frac{1}{\sqrt{2}} \\
	 \frac{1}{2} & -\frac{1}{2} & \frac{1}{\sqrt{2}}
    \end{array}\right) \label{xorder} \\
		&+&
       s_{13} \left( \begin{array}{ccc}
        1 & -1 & e^{-i\delta} \\
        \frac{1}{\sqrt{2}}-\frac{1}{2}e^{+i\delta} & \frac{1}{\sqrt{2}}-\frac{1}{2}e^{+i\delta} & 0 \\
        -\frac{1}{\sqrt{2}}-\frac{1}{2}e^{+i\delta} & -\frac{1}{\sqrt{2}}-\frac{1}{2}e^{+i\delta} & 0
			\end{array} \right)
		+ \mathcal{O}(s_{13}^2). \nonumber
\end{eqnarray}
Suppose $m_1=0$ in an extreme case of normal hierarchical mass ordering. Since only the difference of each Majorana phase relative to overall phase becomes physical, one of the Majorana phases in Eq.(\ref{fulltrans}) can be removed due to vanishing $m_1$. The symmetric neutrino mass matrix in Eq.(\ref{umu}) appears as
\begin{widetext}
	\begin{eqnarray}
		M_\nu = && ~ \tilde{m}_3 \left(
		\begin{array}{ccc}
 		s_{13}^2 e^{-2 \imath\delta} & \frac{1}{\sqrt{2}} s_{13} e^{-\imath\delta} & \frac{1}{\sqrt{2}} s_{13} e^{-\imath\delta} \\
 		\checkmark & \frac{1}{2} & \frac{1}{2} \\
 		\checkmark & \checkmark & \frac{1}{2} \\
	\end{array}\right) ~+ \label{lowmass3} \\
		&& m_2 \left(\begin{array}{ccc}
		 \frac{1}{2}- \sqrt{2} s_{13} ~ &  \frac{1}{2\sqrt{2}}\left(1 - \sqrt{2}s_{13} \right) \left(1 - \sqrt{2}s_{13} + s_{13} e^{\imath\delta}\right) &  -\frac{1}{2\sqrt{2}}\left(1 - \sqrt{2}s_{13} \right) \left(1- \sqrt{2}s_{13} + s_{13} e^{\imath\delta}\right) \\
		 \checkmark & \frac{1}{4} \left(1 - \sqrt{2}s_{13} + s_{13} e^{\imath\delta}\right)^2 & -\frac{1}{4}\left(1 - \sqrt{2}s_{13} \right) \left(1+ \sqrt{2}s_{13} + s_{13} e^{\imath\delta}\right) \\
		\checkmark & \checkmark & \frac{1}{4}\left(1+ \sqrt{2}s_{13} + s_{13} e^{\imath\delta}\right)^2
		\end{array}	\right), \nonumber
	\end{eqnarray}
\end{widetext}
where $\tilde{m}_3 \equiv m_3e^{i\varphi}$ with $\varphi=\varphi_3-\varphi_2$.

The quantities related to the imaginary part of PMNS are Jarlskog invariant $J_{CP}$ and effective electron neutrino mass $<m_{ee}>$. Jarlskog invariant evaluates the magnitude of CP asymmetry from Dirac phase, which can be estimated from oscillation probability in appearance experiments. Thus, from $U$ in Eq.(\ref{xorder}), the size of $J_{CP}$ is expressed in a simple combination of $s_{13}$ and Dirac phase $\delta$ as follows,
	\begin{eqnarray}
	J_{CP}&=&\mathrm{Im}[U_{e1}U_{\tau 3}U_{\tau 1}^*U_{e3}^*] \\
			&=& \frac{1}{4}s_{13}\sin\delta - \frac{1}{2}s_{13}^3\sin\delta + \mathcal{O}(s_{13}^5),
	\end{eqnarray}
The effective electron neutrino mass $<m_{ee}>$ is the only neutrino-dependent factor to determine the amplitude of neutrinoless double beta decay.
	\begin{eqnarray}
	<m_{ee}> &\equiv& |\sum^3_{i=1}U_{ei}^2m_ie^{i\varphi_i}| \\
				&=& m_2\left( \frac{1}{2}-\sqrt{2}s_{13} + s_{13}^2 \right)  \\
				&+& m_3\left( \cos\left(2\delta+\varphi\right)s_{13}^2 +\frac{m_3}{m_2}\sin^2\left(2\delta+\varphi\right)s_{13}^4 \right), \nonumber
	\end{eqnarray}
Since the $<m_{ee}>$ of normal hierarchical masses is far below the sensitivity pursued by on-going neutrinoless double-beta decay experiments, we do not discuss it any longer in this work.

\section{Yukawa Interaction and Seesaw Mechanism}

Canonical model of seesaw mechanism to suppress neutrino masses below 1 eV using heavy neutrino mass scale requires $SU(2)$ Higgs doublet, whose existence was definitely confirmed. The lagrangian density for seesaw mechanism consists of Yukawa couplings of leptons  and lepton-number-violating Majorana mass terms of right-handed heavy neutrinos.
	\begin{eqnarray}
    - \mathcal{L} = H \mathcal{Y}_\ell L_e \bar{e}_R
    + H \mathcal{Y}_\nu L_e \bar{N}_R + \frac{1}{2} M_R N_R N_R,
	\label{lagrangian}
	\end{eqnarray}
The minimal model that assumes CP violation from heavy neutrino decay through Yukawa coupling requires at least two right-handed neutrinos. We consider the following particle contents: $N_R=(N_1, N_2)$ in the basis mass matrix $M_R$ is diagonal, and $\nu_l = (\nu_e, \nu_\mu, \nu_\tau)$ in the basis Yukawa coupling of charged leptons $\mathcal{Y}_\ell$ is diagonal. The $\mathcal{L}$ consists of a $3 \times 3$ matrix $\mathcal{Y}_\ell$, a $3 \times
2$ matrix $\mathcal{Y}_\nu$ and a $2 \times 2$ matrix $M_R$, which naturally result in zero mass for one of light neutrinos through the seesaw mechanism\cite{Gell-Mann:vs}, $M_\nu = - v^2 \mathcal{Y}_\nu M_R^{-1} \mathcal{Y}_\nu^T$ in top-down approach.

In bottom-up approach, it is possible to trace the matrix $\mathcal{Y}_\nu$ in terms of light neutrino masses and mixing angles using the seesaw mechanism in opposite direction.
When the $3 \times 2$ Yukawa matrix is given by
    \begin{equation}
        \mathcal{Y}_\nu \equiv
    \left(
        \begin{array}{cc}
        y_{11} & y_{12} \\
        y_{21} & y_{22} \\
        y_{31} & y_{32}
        \end{array} \right),
    \label{dirac}
    \end{equation}
the Yukawa couplings are imbedded in symmetric neutrino mass matrix in the following way:
    \begin{eqnarray}
        M_\nu &=& \frac{v^2}{M_1}
    \left(\begin{array}{ccc}
        y_{11}^2 & y_{11}y_{21} & y_{11}y_{31} \\
        \checkmark & y_{21}^2 & y_{21} y_{31} \\
        \checkmark & \checkmark & y_{31}^2
    \end{array}\right)  \nonumber \\
	&+& \frac{v^2}{M_2}
    \left(\begin{array}{ccc}
        y_{12}^2 & y_{12} y_{22} & y_{12} y_{32} \\
        \checkmark & y_{22}^2 & y_{22} y_{32} \\
        \checkmark & \checkmark & y_{32}^2
    \end{array}\right)\label{m2seesaw},
    \end{eqnarray}
where $M_1$ and $M_2$ are the masses of $N_1$ and $N_2$, respectively, and $v$ is the vacuum expectation value of $H$.

In comparison of the $M_\nu$ in terms of low-energy physical parameters Eq.(\ref{lowmass3}) with the $M_\nu$ in terms of Yukawa couplings Eq.(\ref{m2seesaw}), we obtained the following relations;
    \begin{eqnarray}
		&& \left( \begin{array}{c} y_{11} \\ y_{21} \\ y_{31}  \end{array} \right)=
			\sqrt{\frac{m_3}{m_2}}e^{\imath\varphi/2} \left( \begin{array}{l} s_{13} e^{-\imath\delta} \\
			\frac{1}{\sqrt{2}} \\
			\frac{1}{\sqrt{2}} \end{array} \right) \label{yuk11} \\
		&& \left( \begin{array}{c} y_{12} \\ y_{22} \\ y_{32}  \end{array} \right)=
			\sqrt{\frac{M_2}{M_1}} \left( \begin{array}{l}
			\frac{1}{\sqrt{2}} - s_{13}  \\
			\frac{1}{\sqrt{2}} \left( \frac{1}{\sqrt{2}}-s_{13} + \frac{1}{\sqrt{2}}s_{13} e^{\imath\delta}\right) \\
			-\frac{1}{\sqrt{2}} \left( \frac{1}{\sqrt{2}}+s_{13} + \frac{1}{\sqrt{2}}s_{13} e^{\imath\delta}\right)\end{array} \right). \label{yuk12}
    \end{eqnarray}
All Yakawa couplings are determined in terms of 5 physical values: $s_3, \delta, \varphi,\sqrt{m_3/m_2}$, and $\sqrt{M_2/M_1}$, implying there can be a way to test the model and constrain the scale of seesaw mechanism from experimental measurements.

Once we have the neutrino Yukawa matrix with the couplings in Eq.(\ref{yuk11}) and Eq.(\ref{yuk12}), we can calculate the
magnitude of CP asymmetry $\epsilon_i$ in decays of heavy Majorana
neutrinos \cite{Kolb:qa,Luty:un},
\begin{eqnarray}
\epsilon_i
        &=&\frac{\Gamma (N_i \to \ell H)
            - \Gamma (N_i \to \bar{\ell} H^*)}
        {\Gamma (N_i \to \ell H)
            + \Gamma (N_i \to \bar{\ell} H^*)},
\label{aacp}
\end{eqnarray} %
where $i$ denotes a generation. If the scale of $M_1$ is far below $M_2$, the CP asymmetry is solely obtained from the decay of $M_1$ \cite{Kolb:qa,Luty:un}. Hence, the asymmetry is replaced by just $\epsilon_1$ whose magnitude is given by
\begin{eqnarray}
    \epsilon_1 &=& \frac{1}{8\pi}
            \frac{{\rm Im}\left[(\mathcal{Y}_\nu^\dagger \mathcal{Y}_\nu)_{12}^2\right]
            } {(\mathcal{Y}_\nu^\dagger \mathcal{Y}_\nu)_{11}}
            f\left(\frac{M_2}{M_1}\right)\;, \label{cp1}
\end{eqnarray}
where $f\left(M_2/M_1\right)$ represents loop contribution to the
decay width from vertex and self energy.
In the Standard Model, it is given by
\begin{equation}
f(x) = x\left[1-(1+x^2)\ln \frac{1+x^2}{x^2} +
\frac{1}{1-x^2}\right],
\end{equation}
which can be approximated to $3/2x$ for sufficiently large $x$. Then the asymmetry $\epsilon_1$ in Eq.(\ref{cp1}) can be written as
    \begin{eqnarray}
		  && \epsilon_1 = \frac{3}{16\pi}\Delta_1 \frac{M_1}{M_2},  \label{finalCP}
	 \end{eqnarray}
for $M_1 \ll M_2$. The $\Delta_1$ factor that depends on Yukawa couplings is
	 \begin{eqnarray}
        && \Delta_1 \equiv \frac{{\rm Im}\left[(\mathcal{Y}_\nu^\dagger \mathcal{Y}_\nu)_{12}^2\right]}
            {(\mathcal{Y}_\nu^\dagger \mathcal{Y}_\nu)_{11}} \label{delta1}  \\
			&& = \frac{Im \left[( y_{11}^* y_{12} + y_{21}^* y_{22} + y_{31}^* y_{32})^2\right]}
        {|y_{11}|^2 + |y_{21}|^2 +|y_{31}|^2 },
		  \nonumber  \\
			&& = \frac{\mu^2 s_{13}^2}{1+s_{13}^2}
			\left( \sqrt{2}\sin\left(\varphi-\delta\right)-\frac{1}{2}\sin\left(\varphi-2\delta\right)-\sin\varphi \right), \nonumber \end{eqnarray}
with
	\begin{eqnarray}
	\mu^2=\frac{M_2}{M_1}.
	\end{eqnarray}
Then, the ratio of heavy Majorana neutrino masses is eliminated in $\epsilon_1$, so that one can have $\epsilon_1=\epsilon_1(\delta, \varphi, s_{13})$ regardless of the relative mass scale of heavy neutrinos.

\section{Lepton Asymmetry}
The baryon density of our universe $\Omega_B h^2 = 0.02240$ from nine-year Wilkinson Microwave Anisotropy Probe(WMAP) data
indicates the observed baryon asymmetry in the
Universe\cite{WMAP1},
    \begin{equation}
    \eta_B^{CMB}=
    \frac{n_B-n_{\bar{B}}}{n_\gamma}= 6.5 \times 10^{-10},
    \label{baryon}
    \end{equation}
where $n_B, n_{\bar{B}}$ and $n_\gamma$ are number density of
baryon, anti-baryon and photon, respectively. The baryon asymmetry
Eq.(\ref{baryon}) can be rephrased by
    \begin{equation}
    Y_B = \frac{n_B-n_{\bar{B}}}{s} \simeq
    9.8\times 10^{-11}. \label{cosmo}
    \end{equation}
with the entropy density $s$ in order to take into account
the number density with respect to a co-moving volume element.
The baryon asymmetry produced through sphaleron process is related to the
lepton asymmetry \cite{Harvey:1990qw} by $Y_B = \frac{a}{a-1} Y_L$
with $ a \equiv (8 N_F + 4 N_H) / (22 N_F + 13 N_H)$. For the Standard Model(SM), $a=28/79$ with three generations of fermions and a single Higgs doublet. The lower bound of $Y_B=9.8\times 10^{-11}$ in Eq.(\ref{cosmo}) can be replaced by $Y_L=1.8 \times 10^{-10}$.
The generation of a lepton asymmetry requires the CP-asymmetry and
out-of-equilibrium condition. The $Y_L$ is explicitly
parameterized by two factors, $\epsilon$, the size of CP
asymmetry, and $\kappa$, the dilution factor from washout process.
\begin{eqnarray}
Y_L = \frac{(n_L - n_{\overline{L}})}{s} = \kappa
\frac{\epsilon_i}{ g^*} \label{aalepto}
\end{eqnarray}
where $g^*\simeq 110$ is the number of relativistic degree of
freedom.

The $\kappa$ in Eq.(\ref{aalepto}) is determined by solving the
full Boltzmann equations. The $\kappa$ can be simply parameterized
in terms of $K$, which is defined as the ratio of $\Gamma_1$ the tree-level
decay width of $N_1$ to $H$ the Hubble parameter at
temperature $M_1$, $\Gamma_1 / H$. The condition $K<1$ describes
processes out of thermal equilibrium and $\kappa<1$ describes
washout effect\cite{Harvey:1990qw},
\begin{eqnarray}
    \kappa = \frac{0.3}{K \left(\ln K \right)^{0.6}}
    ~&~\rm{for}~&~ 10 \lesssim K \lesssim 10^6, \label{largek} \\
        \kappa = \frac{1}{2 \sqrt{K^2+9}}
        ~&~\rm{for}~&~ 0 \lesssim K \lesssim 10. \label{smallk}
\end{eqnarray}
The decay width of $N_1$ by the Yukawa interaction at tree level
and Hubble parameter in terms of temperature $T$ and the Planck
scale $M_{pl}$ are $\Gamma_1 =  (\mathcal{Y}_\nu^\dagger \mathcal{Y}_\nu)_{11}
M_1 / (8 \pi) $ and $H = 1.66 g^{1/2}_* T^2 / M_{pl}$,
respectively. At temperature $T = M_1$, the ratio $K$ is
    \begin{eqnarray}
    K = \frac{M_{pl}}{1.66 \sqrt{g^*}(8 \pi)}
    \frac{(\mathcal{Y}_\nu^\dagger \mathcal{Y}_\nu)_{11}}{M_1}, \label{yukawak}
    \end{eqnarray}
which reduces to
    \begin{eqnarray}
    K = \frac{M_{pl}/M_1}{1.66 \sqrt{g^*}(8 \pi)} \frac{m_3}{m_2},
    \label{kay}
    \end{eqnarray}
since $(\mathcal{Y}_\nu^\dagger
\mathcal{Y}_\nu)_{11}=\sum|y_{i1}|^2 = m_3/m_2 $ up to the leading order of $s_{13}$ from Eq.(\ref{yuk11}),
Then the dilution factor $\kappa$ depends on the relative scale of $M_{pl}/M_1$ and the ratio of light masses $m_3/m_2$, so that
    \begin{eqnarray}
        Y_L =
        \frac{1}{g^*}~\kappa~(\frac{M_{pl}}{M_1},\frac{m_3}{m_2})
        ~\epsilon_1(\delta,\varphi,s_{13}), \label{YL_inputs}
    \end{eqnarray}
which implies that the thermal effect is a matter of mass scales while the CP asymmetry is a matter of complex mixing of particles.

\begin{figure}
\resizebox{90mm}{!}{\includegraphics[width=0.75\textwidth]{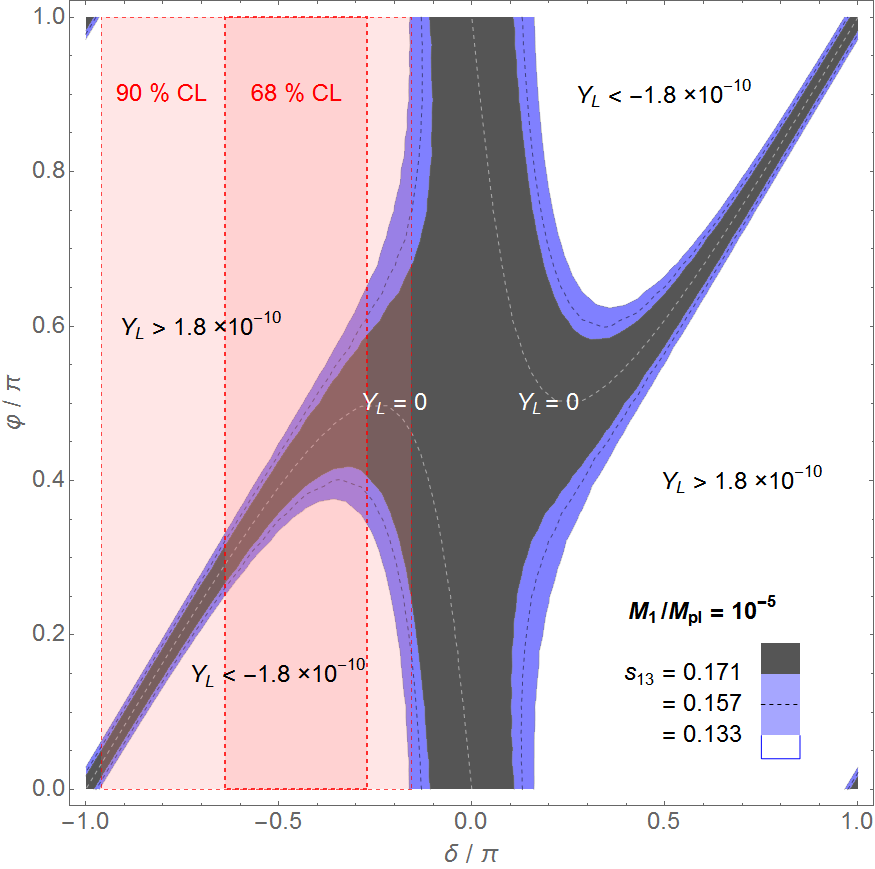}}
\caption{\label{fig2:fig_theta13_contours}
Contours of $Y_L = \pm 1.8 \times 10^{-10}$, where $s_{13}$ varies within $3\sigma$ range in global analysis. The dark shade in the center indicates the range excluded by the cosmological bound. The boundaries between the gray shade and the blue bands represent $Y_L$ corresponding to the value $s_{13}=0.171$. The dashed curves and the outmost curves in the blue bands represents $Y_L$ for $s_{13}=0.157$ and $0.133$, respectively. Red shades are the $\delta_{CP}$ ranges of T2K at 68\% CL and 90\% CL.}
\end{figure}

\begin{figure}
\resizebox{90mm}{!}{\includegraphics[width=0.75\textwidth]{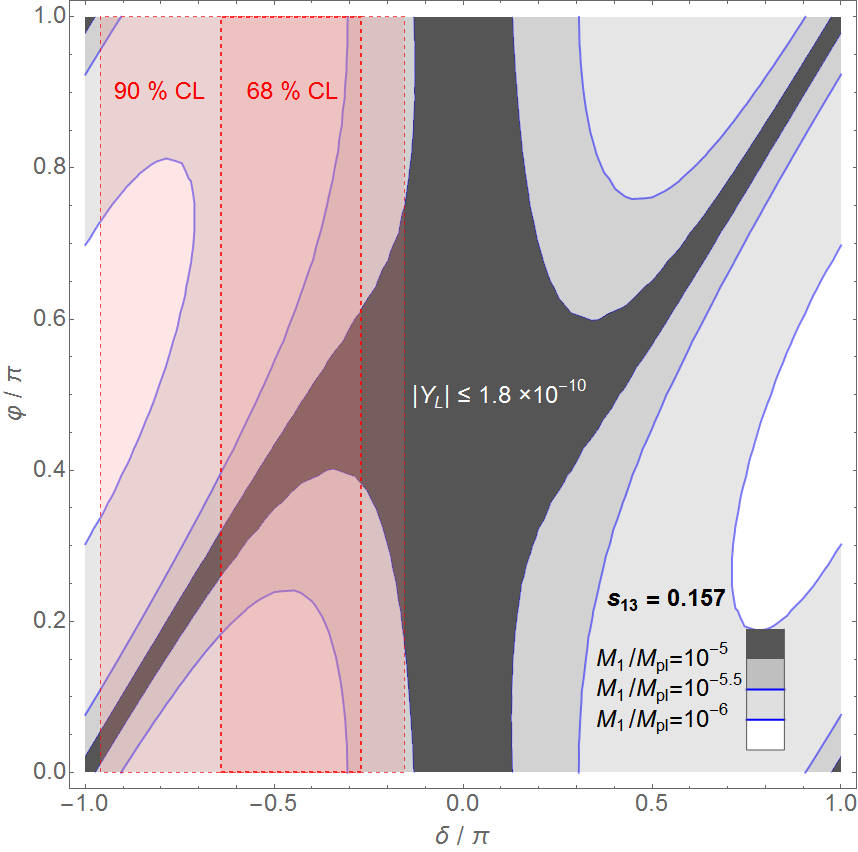}}
\caption{\label{fig3:fig_seesaw_scale}
Contours of $Y_L = \pm 1.8 \times 10^{-10}$ depending on $M_1/M_{\mathrm{Pl}}$. The boundaries of dark shade are obtained from $M_1/M_{\mathrm{Pl}}=10^{-5}$. The curves from $M_1/M_{\mathrm{Pl}}=10^{-5.5}$ and $10^{-6}$ extend the excluded region so as to narrow down $|Y_L|>1.8\times 10^{-10}$ areas.  }
\end{figure}

In this framework of bottom-up approach, some of low energy quantities can be fixed by best fit in global analysis,
	\begin{eqnarray}
	\frac{m_2}{m_3} &=& \sqrt{\frac{\Delta m_{21}^2}{\Delta m_{31}^2}} ~\approx~ 0.176, \\
	s_{13} &=& \sin\theta_{13} ~\approx~ 0.157, \label{bestfit}
	\end{eqnarray}
where their $3\sigma$-ranges are given by $(0.164~-~0.192)$ and by $(0.133~-~0.171)$, respectively. The curves in Fig. \ref{fig2:fig_theta13_contours} present the contours of lepton asymmetry $Y_L=\pm 1.8\times 10^{-10}$ for $m_2/m_3$ given in Eq.(\ref{bestfit}) and a selected scale of $M_1/M_{pl}$. The region below the size of $Y_L$ derived from the cosmological bound $Y_B$ is shaded, excluding the corresponding combination of $\varphi$ and $\delta$ . Both signs of asymmetry are taken into account. It is clear that the amount of lepton asymmetry cannot be reached to the cosmological bound without Dirac phase, although they are not proportional to each other. In this model, $\delta=0$ is excluded regardless other variables in Eq.(\ref{YL_inputs}) as long as their values are considered within phenomenologically allowed ranges.
It is also shown that the variation of $\sin\theta_{13}$ within $3\sigma$ range does not draw a remarkable change in lepton asymmetry $Y_L$, and neither does the variation of $m_2/m_3$ in $3\sigma$. The changes in $Y_L$ affected by $m_2/m_3$ is barely visible. The experiments of long-baseline oscillation began obtaining results on Dirac CP phase. The red shades in Fig.\ref{fig2:fig_theta13_contours} and Fig.\ref{fig3:fig_seesaw_scale} present the ranges of $\delta$ at 68\% and 90\% CL obtained by T2K with best fit $-0.5\pi$. The CP conservation is ruled out at 90\% CL.


The amount of lepton asymmetry in Eq.(\ref{aalepto}) is now given
as a function of $\delta$ and $\varphi$ as well as $M_1/M_{pl}$ such as
    \begin{eqnarray}
        Y_L =
        \frac{1}{g^*}\kappa\left(\frac{M_1}{M_{pl}}\right)
        \epsilon_1\left(\delta,\varphi\right), \label{finalYL}
    \end{eqnarray}
because the variation in $m_2/m_3$ and that in $s_{13}$ do barely affect $Y_L$.
It turns out that the washout effect of asymmetry is mainly affected by the lightest mass of heavy Majorana neutrino $M_1$, in other word, the scale of seesaw mechanism. It appears as the ratio to the planck scale $M_1/M_{pl}$. According to Eq.(\ref{smallk}), there is an upper bound 17\% to the dilution factor $\kappa$ no matter how strong the out-of-equilibrium condition is.  The relation of $K$ with $M_1$ in Eq.(\ref{yukawak}) implied that the lower $M_1$ scale becomes, the more asymmetry is washed out, as shown in Fig.\ref{fig3:fig_seesaw_scale}. The areas enclosed by different contours indicate the cosmological bound derived from different scales of $M_1$. As the seesaw scale $M_1$ decreases, the region for sufficient asymmetry becomes narrower. For example, the scale $M_1$ below $10^{-6}M_{pl}$ is not compatible with the Dirac CP phase within T2K 68\% CL.

\section{Conclusion}

In a minimal seesaw model with two right-handed neutrinos, the lepton asymmetry $Y_L$ for Baryogenesis can be probed by low-energy phenomenology. For sufficient $Y_L$, the model requires non-zero $\delta$. Although the size of $Y_L$ depends on the values of $m_2/m_3$ and $s_{13}$, its variations within current uncertainties of those parameters are almost invisible. Once mixing angles are fixed, the asymmetry is described by seesaw-scale factor and by experimentally measurable factor, as shown in Eq.(\ref{finalYL}). One can expect that the precise measurement of $\delta$ in future experiments can constrain the right-handed neutrino mass for seesaw mechanism.


\appendix

\nocite{*}


\end{document}